\documentclass[reprint,twocolumn,amsmath,superscriptaddress,showpacs,aps,prl]{revtex4-1}
\pdfoutput=1

\usepackage{graphicx}% Include figure files
\usepackage{dcolumn}% Align table columns on decimal point
\usepackage{bm}% bold math
\usepackage{epsfig}
\usepackage{amsmath}
\usepackage{natbib}
\usepackage{color}

\begin{document}
\title{Amplitude \textquoteleft Higgs' mode in 2H-NbSe$_2$ Superconductor}

\author{M.-A. M\'easson}
\email[E-mail: ]{marie-aude.measson@univ-paris-diderot.fr}
\affiliation{Laboratoire Mat\'eriaux et Ph\'enom\`{e}nes Quantiques (UMR 7162 CNRS),
Universit\'e Paris Diderot-Paris 7, Bat. Condorcet, 75205 Paris Cedex 13, France.}
\author{Y. Gallais}
\affiliation{Laboratoire Mat\'eriaux et Ph\'enom\`{e}nes Quantiques (UMR 7162 CNRS),
Universit\'e Paris Diderot-Paris 7, Bat. Condorcet, 75205 Paris Cedex 13, France.}
\author{M. Cazayous}
\affiliation{Laboratoire Mat\'eriaux et Ph\'enom\`{e}nes Quantiques (UMR 7162 CNRS),
Universit\'e Paris Diderot-Paris 7, Bat. Condorcet, 75205 Paris Cedex 13, France.}
\author{B. Clair}
\affiliation{Laboratoire Mat\'eriaux et Ph\'enom\`{e}nes Quantiques (UMR 7162 CNRS),
Universit\'e Paris Diderot-Paris 7, Bat. Condorcet, 75205 Paris Cedex 13, France.}
\affiliation{Laboratoire Structure Propri\'{e}t\'{e} et Mod\'{e}lisation des Solides (CNRS UMR 8580),
Ecole Centrale Paris, 92 295 Chatenay Malabry, France.}
\author{P. Rodi\`{e}re}
\affiliation{Institut N\'eel, CNRS-UJF, 25 Av. des Martyrs, 38042 Grenoble, France.}
\author{L. Cario}
\affiliation{Institut des Materiaux Jean Rouxel (IMN), Universite de Nantes - CNRS, 2 rue de la Houssiniere, BP 32229, 44322 Nantes Cedex 03, France.}
\author{A. Sacuto}
\affiliation{Laboratoire Mat\'eriaux et Ph\'enom\`{e}nes Quantiques (UMR 7162 CNRS),
Universit\'e Paris Diderot-Paris 7, Bat. Condorcet, 75205 Paris Cedex 13, France.}

%\date{\today}

\begin{abstract}

We report experimental evidences for the observation of the superconducting amplitude mode, so-called \textquoteleft Higgs' mode in the charge density wave superconductor 2H-NbSe$_2$ using Raman scattering. By comparing 2H-NbSe$_2$ and its iso-structural partner 2H-NbS$_2$ which shows superconductivity but lacks the charge density wave order, we demonstrate that the superconducting mode in 2H-NbSe$_2$ owes its spectral weight to the presence of the coexisting charge density wave order. In addition temperature dependent measurements in 2H-NbSe$_2$ show a full spectral weight transfer from the charge density wave mode to the superconducting mode upon entering the superconducting phase. Both observations are fully consistent with a superconducting amplitude mode or Higgs mode.

\end{abstract}

%78.30.Er 	Solid metals and alloys
%78.30.-j 	Infrared and Raman spectra (
%for vibrational states in crystals  see 63.20.-e
%for Raman spectra of superconductors, see 74.25.nd)

%78.70.Nx 	Neutron inelastic scattering

%63.20.D- 	Phonon states and bands, normal modes, and phonon dispersion
%63.20.dk 	First-principles theory
%63.20.kd 	Phonon-electron interactions

%71.27.+a 	Strongly correlated electron systems; heavy fermions

%78.70.Ck: x-ray scattering
%71.45.Lr 	Charge-density-wave systems

%74.70.Ad 	Metals; alloys and binary compounds (including A15, MgB2, etc.)


\maketitle

While the quest for the Higgs boson in particle physics is reaching its goal and its prediction has been rewarded by the Nobel prize, there is growing interest in the search for an analogous excitation in quantum many body systems where the Higgs boson manifests itself as a fundamental collective mode \cite{Varma2002,Nambu,Podolsky}.
When a spontaneous breaking of a continuous symmetry takes place collective excitations of the order parameter emerge: they are the massless Nambu-Goldstone phase modes \cite{Anderson} and the massive amplitude Higgs mode\cite{Higgs1964}. In quantum many body systems, the Higgs mode was recently identified in ultra-cold 2D bosonic $^{87}$Rb atoms in optical lattice \cite{endres} and reported in a the dimer antiferromagnet TlCuCl$_3$ \cite{Ruegg2008}. Very recently, it has been unveiled in the superconducting Nb$_{1-x}$Ti$_x$N films by using terahertz pump probe spectroscopy\cite{Matsunaga2013}. The existence of a Higgs mode was proposed more than thirty years ago in the bulk charge density wave (CDW) superconductor (SC) 2H-NbSe$_2$ \cite{Varma1981,Varma1982}, where a superconducting amplitude mode, or Higgs mode, can be unraveled via its coupling to the coexisting charge density wave mode. Despite this prediction after the first experimental observation by Raman scattering\cite{Sooryakumar80,Sooryakumar81}, unambiguous proofs of its Higgs type nature have remained elusive up to now\cite{Balseiro,Lei,Tutto}.

As it does not carry any spin or charge, in principle, the amplitude mode of the superconducting order parameter, or the Higgs mode, does not couple directly to any external probe. However, when superconductivity coexists with a charge density wave order, the amplitude mode of the CDW order couples to the Higgs mode by modulating the density of states at the Fermi level, thus "shaking" the SC condensate by modulating the amplitude of the superconducting order parameter. This allows the indirect detection of the 'Higgs' mode by spectroscopic probes \cite{Varma1981,Varma1982}. Experimentally, the Higgs mode becomes active by removing spectral weight from the CDW amplitude mode upon entering the SC state. The requisite of a coexisting CDW mode and the observation of a transfer of spectral weight from the CDW amplitude mode to the Higgs mode in the SC state can thus be considered as key predictions of the Higgs mode scenario. \par
Raman inelastic light scattering experiments allow access to symmetry dependent collective excitations of both CDW and SC orders. Crucially, because the amplitude mode of the CDW order is Raman active \cite{Tsang}, it is ideally suited for the detection of the Higgs mode in the SC state. The Raman response of 2H-NbSe$_{2}$ has been extensively studied~ \cite{Tsang,Wang,Pereira,Wu,Mialitsin}, including at low temperature, in the superconducting phase \cite{Sooryakumar80,Sooryakumar81}. On the contrary, 2H-NbS$_2$ has been little investigated by Raman spectroscopy~\cite{Nakashima,McMullan85} and no attempt to perform measurements in the superconducting state has been performed. Because 2H-NbS$_2$ does not show any CDW order, its comparison with 2H-NbSe$_2$ provides a stringent test for the Higgs mode scenario.

We show here, by comparison of quantitative Raman scattering measurements in 2H-NbSe$_2$ and 2H-NbS$_2$, that the narrow and intense superconducting mode in 2H-NbSe$_2$ cannot be a simple Cooper-pair breaking mode. In 2H-NbS$_2$, only a much weaker Cooper pair breaking peak is observable. Clearly, the coexisting CDW mode is necessary for the observation of the intense superconducting mode in 2H-NbSe$_2$. We also report an almost perfect transfer of spectral weight from the CDW mode to the SC mode from $T_c$ down to 2~K in 2H-NbSe$_2$ in A$_{1g}$ and E$_{2g}$ symmetries, showing that the SC mode draws all its Raman intensity from the coexisting CDW mode. Both experimental observations are new strong evidences for the Higgs mode scenario in 2H-NbSe$_2$.

\par
The dichalcogenide 2H-NbSe$_2$ offers the unique feature of exhibiting charge density wave and superconducting ordering temperatures of the same order of magnitude ($T_c$=7.1~K and $T_{CDW}$=33~K respectively). This property allows an efficient coupling between the CDW mode and the amplitude mode of the SC order parameter, a pre-requisite for the detection of the amplitude SC mode or Higgs mode. Whereas 2H-NbSe$_2$ exhibits coexisting SC and CDW orders, its isostructural and isoelectronic partner 2H-NbS$_2$ presents only the superconducting state with comparable $T_c$ (6.05~K). Both compounds have similar superconducting properties which have been interpreted as either multiple SC gaps or a single anisotropic SC gap \cite{Boaknin, GuillamonSe, Kiss, Diener, GuillamonS, kacmarcik}.
Moreover the Fermi surface of both compounds seems to differ only by the absence of the pancake-like sheet at the $\Gamma$ point in NbS$_2$ \cite{Takahashi} whereas the two cylindrical Fermi-surface sheets centered at the $\Gamma$ and $K$ points are present in both compounds \cite{Takahashi,Kiss,Borisenko}. Recent inelastic x-ray scattering have shown that 2H-NbS$_2$ is on the verge of CDW order with a softening of the phonon modes along $\Gamma M$ at the same wave vector than the soft mode which drives the CDW transition in 2H-NbSe$_2$. Most probably, in 2H-NbS$_2$, CDW does not occur only due to large anharmonic effects\cite{Leroux2012}.

The 2H-NbS$_2$ single crystal was grown using the vapor transport growth technique with a large sulfur excess \cite{Fisher1980}. It was characterized by IXS \cite{Leroux2012}, STM \cite{GuillamonS}, specific heat \cite{kacmarcik} and penetration depth \cite{Diener} measurements. All measurements indicate a high quality of this single crystal, with a sharp superconducting transition measured by specific heat of $0.05\ast \!T_c$. The 2H-NbSe$_2$ single crystal was synthesized using an iodine-vapour transport method. It has a residual resistivity ratio of 39 and the transition to the charge density wave order is clearly visible at 33K by resistivity measurement, confirming its good quality.

Raman scattering investigation has been carried out at low temperature down to 2~K. Single crystals of 2H-NbSe$_2$ and 2H-NbS$_2$, freshly cleaved, were cooled down in a $^4He$ pumped cryostat during the same experiment to be able to quantitatively compare the intensity of the Raman signals.
The samples were kept in $^4He$ exchange gas allowing a reduced laser heating \cite{sup1}. The reported temperatures include the laser heating. We have taken special care to avoid and check that the samples were not immersed in $^4He$ liquid, to avoid the Raman signature of superfluid $^4He$ at 13~cm$^{-1}$.

Raman scattering measurements have been performed in a quasi-backscattering geometry with a triple spectrometer Jobin Yvon T64000 equipped with a liquid-nitrogen-cooled CCD detector. We have used the 514~nm excitation line from a Ar$^+$-Kr$^+$ mixed gas laser. The polarization for the incoming and outgoing light are in the (ab) plane of the sample. In the 2H-dichalcogenides, four Raman active symmetries are expected, one A$_{1g}$, one E$_{1g}$, and two E$_{2g}$. In our configuration ($\overrightarrow{E}\in(ab)$), we detect the A$_{1g}$ and the two E$_{2g}$. In parallel and crossed polarizations we select the A$_{1g}$+E$_{2g}$ and the E$_{2g}$ modes, respectively. By scaling the E$_{2g}$ phonon mode at about 250~cm$^{-1}$, the A$_{1g}$ pure symmetry scattering is deduced by the subtraction of the E$_{2g}$ signal from the A$_{1g}$+E$_{2g}$ signal.

\begin{figure}
\includegraphics*[width=8cm]{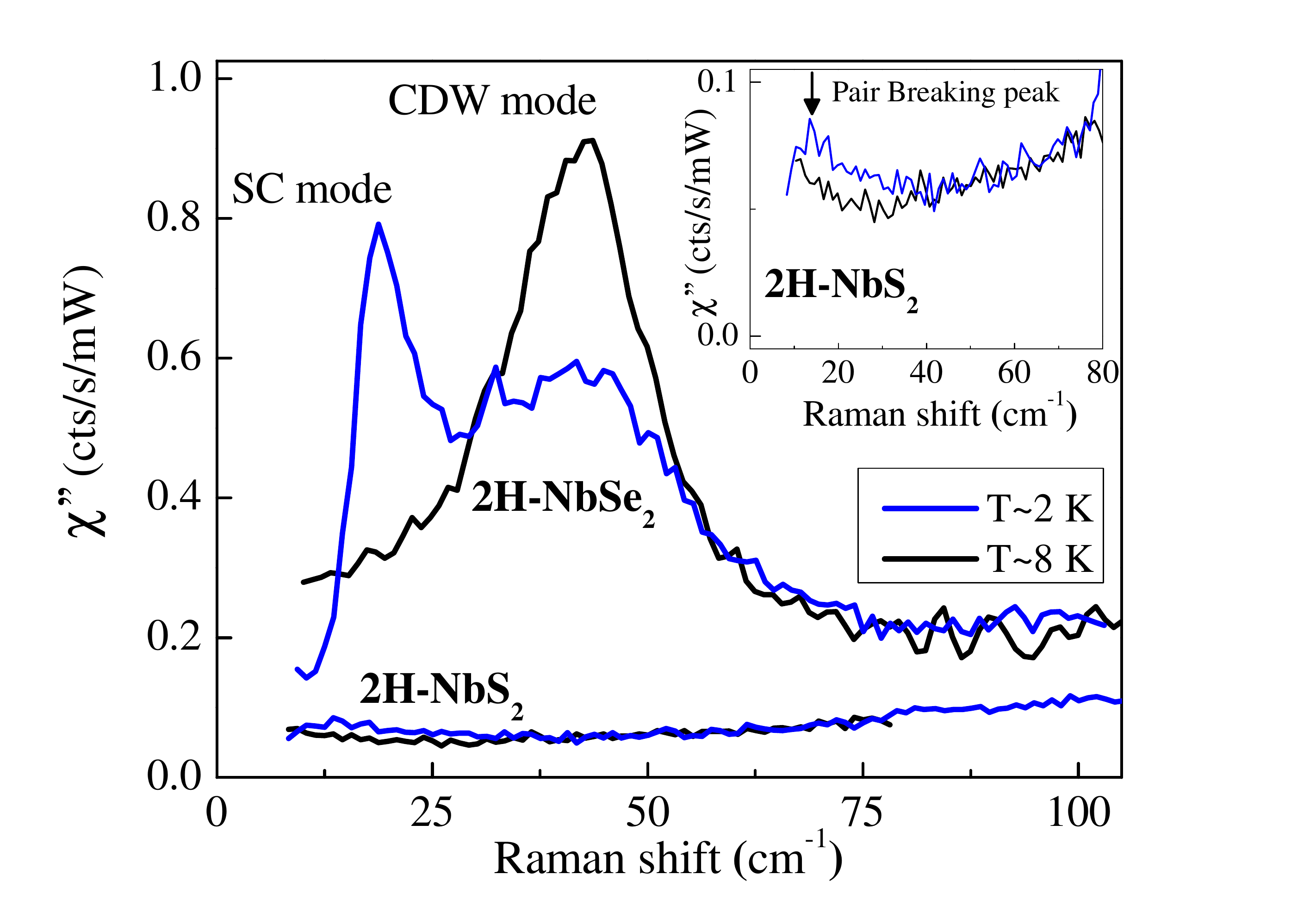}
\caption{\label{Comp} (color online) Quantitative Raman susceptibility for 2H-NbSe$_2$ and 2H-NbS$_2$ just above $T_c$ and below $T_c$ at 2~K in A$_{1g}$ + E$_{2g}$ symmetry. SC mode: amplitude mode or 'Higgs' mode of the superconducting state of 2H-NbSe$_2$. CDW mode: Amplitude mode of the charge density wave order of 2H-NbSe$_2$. The inset is a zoom on the superconducting Cooper pair breaking peak in NbS$_2$. It is much less intense than the SC mode in NbSe$_2$.}
\end{figure}

Figure~\ref{Comp} shows the Raman spectra in the superconducting state, at 2~K, and just above $T_c$ in 2H-NbSe$_2$ and 2H-NbS$_2$ and in the A$_{1g}$+E$_{2g}$ symmetry channel. In NbSe$_2$, the peak at $\sim40~cm^{-1}$, labeled CDW mode, has been attributed to the amplitude mode of the CDW order \cite{Tsang}. A narrow and intense peak at $\sim19~cm^{-1}$, labeled SC mode, develops below $T_c$, thus clearly relating its origin to the superconducting state \cite{Sooryakumar81, Sooryakumar80}.
Both modes are present in the A$_{1g}$ and the E$_{2g}$ symmetry channels. As shown in the inset of Fig.~\ref{Comp}, in NbS$_2$, a small peak develops below $T_c$ at 14~cm$^{-1}$. Contrary to the SC mode in NbSe$_2$ it is broad, spreading up to 30~cm$^{-1}$. Its energy matches the largest SC gap measured by STM \cite{GuillamonS}. In this compound with a single electronic SC order, it is most likely due to a Cooper pair-breaking excitation \cite{Abrikosov1961,Abrikosov1973,Klein1984,DevereauxRMP}.

\begin{figure}[!ht]
\includegraphics*[width=8cm]{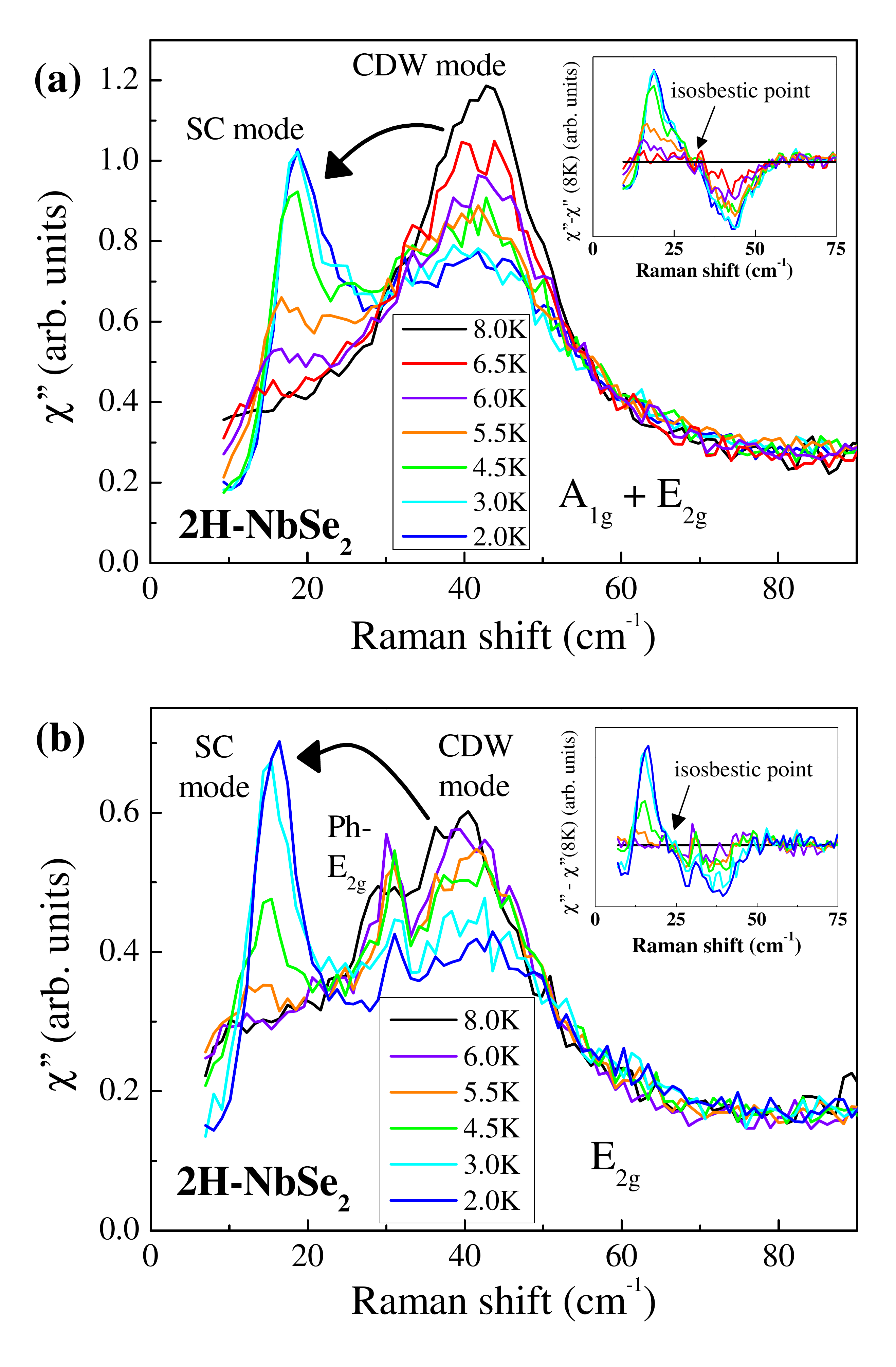}
\caption{\label{RT}(color online) Temperature dependence of the Raman susceptibility of the charge density wave mode and the superconducting mode in 2H-NbSe$_2$. (a) in the A$_{1g}$+E$_{2g}$ symmetry (b) in the E$_{2g}$ symmetry. Ph-E$_{2g}$ is the interlayer phonon mode. The black arrows depict the spectral weight transfer from the CDW mode to the SC mode upon entering the SC state. The insets show the Raman spectra subtracted from the one measured at 8~K above $T_c$ and the arrows point to the isosbestic points, at 29~cm$^{-1}$ and 24~cm$^{-1}$ in the A$_{1g}$+E$_{2g}$ and the E$_{2g}$ symmetries, respectively.}
\end{figure}

By comparing the quantitative Raman signals, it is clear that the intensities of both SC excitations differ drastically. By subtracting the background obtained just above $T_c$, the Cooper pair-breaking excitation in NbS$_2$ is more than 20 times smaller than the SC mode in NbSe$_2$. We conclude that the intense and narrow SC mode in NbSe$_2$ owes its Raman intensity to the presence of the coexisting CDW order.
The huge difference between the intensity of the SC mode in the two compounds clearly demonstrates that in NbSe$_2$, the SC mode cannot be a simple Cooper-pair breaking peak like in NbS$_2$. On the other hand, our observation is fully consistent with the Higgs mode scenario \cite{Varma1981} whereby, in NbSe$_2$, the SC amplitude mode, or Higgs mode, is activated via its coupling to the CDW amplitude mode. In this scheme, the absence of CDW order in NbS$_2$ makes the Higgs mode unobservable leaving only a much weaker Cooper pair breaking peak as observed experimentally.

Additional evidence for the Higgs mode scenario can be obtained from the temperature dependence of the SC and CDW modes of NbSe$_2$. Figure~\ref{RT} displays the Raman response of the electronic modes at various temperature from 2~K up to 8~K above $T_c$ and in A$_{1g}$+E$_{2g}$ and E$_{2g}$ symmetry channels.
At 2~K, the SC modes are at 18.8$\pm 0.5$~cm$^{-1}$ and 16.2$\pm 0.5$~cm$^{-1}$ in A$_{1g}$+E$_{2g}$ and E$_{2g}$ symmetry, respectively; it is extracted to be at 19.2$\pm 0.5$~cm$^{-1}$ in A$_{1g}$ symmetry, consistently with Sooryakumar et al. \cite{Sooryakumar80}. The SC mode softens with increasing temperature whereas the position of the CDW mode stays almost constant in A$_{1g}$+E$_{2g}$ symmetry and slightly shifts to lower energy ($\sim$~3~cm$^{-1}$) in E$_{2g}$ symmetry. Crucially, the SC and CDW modes develop in opposite way: when superconductivity is gradually destroyed, the SC mode intensity collapses while the CDW mode intensity recovers. Quantitative analysis of the spectral weight transfer as a function of temperature in both A$_{1g}$ and E$_{2g}$ symmetries is displayed in Fig.~\ref{IvsT}. The pure A$_{1g}$ symmetry response is shown in the supplemental material\cite{sup2}. The spectral weight is obtained by integrating the imaginary part of the susceptibility $S=\int_{\omega_1}^{\omega_2} \! \chi''(\omega) \, \mathrm{d}\omega.$. The spread of the spectral weight is defined as the deviation from the average value $<S>$, i.e. $\frac{S-<S>}{<S>}$. $<S>$ is defined as the average on the 7 measured spectra (see Fig.~\ref{RT}). The isosbestic point is a phenomenological definition of the energy where all the spectra intersect each others.
As shown in Fig.~\ref{IvsT}a,b, in both symmetries, the spectral weight of the CDW mode and the SC mode evolve in opposite way with temperature. While the spectral weight is transferred at a noticeably faster rate in A$_{1g}$ symmetry than in E$_{2g}$ symmetry below $T_c$, in both symmetries the total spectral weight of cumulative modes is almost constant, with a spread at only $\pm~4~\%$ as reported in Fig.~\ref{IvsT}c.

\begin{figure}[!ht]
\includegraphics*[width=8cm]{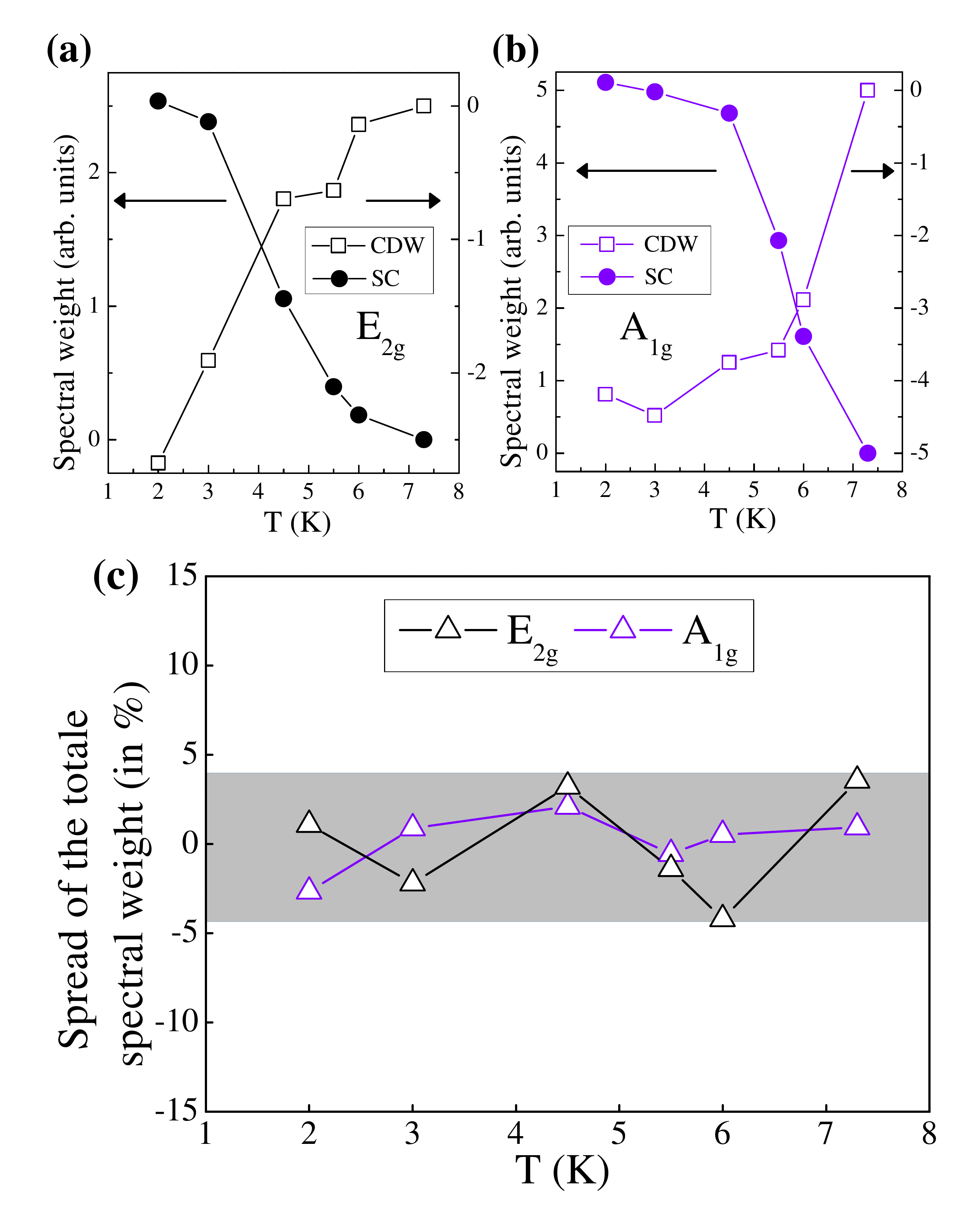}
\caption{\label{IvsT}(color online) (a) and (b) Spectral weight of the SC and CDW modes of 2H-NbSe$_2$ versus temperature, calculated on the subtracted data (Cf. insets of Fig.~\ref{RT}) from 8~cm$^{-1}$ up to the isosbestic point and from the isosbestic point up to 60~cm$^{-1}$ for the SC and the CDW modes, respectively. (a) E$_{2g}$ symmetry (b) A$_{1g}$ symmetry. (c) Percentage of spread of the total spectral weight (CDW mode + SC mode) calculated between 8 and 60~cm$^{-1}$ in E$_{2g}$ and A$_{1g}$ symmetries versus temperature.}
\end{figure}

The almost perfect transfer of spectral weight from the CDW mode to the SC mode below $T_c$ is a strong evidence that the SC mode draws all its Raman intensity from the coexisting CDW mode, in full agreement with the Higgs mode scenario. We note here that a similar transfer of spectral weight was observed by Sooryakumar et al. \cite{Sooryakumar80,Sooryakumar81} upon applying a magnetic field. They calculated the total spectral weight in the A$_{1g}$ symmetry as $S=\int_{\omega_1}^{\omega_2} \! \chi''(\omega)\ast\omega \, \mathrm{d}\omega.$ and found it to be constant within $\pm~7\%$. In our case however the SC and CDW orders are tuned using temperature, without any ambiguous consideration due to the physics of vortex with spatially varying SC order parameter. It should also be emphasized that, our spectral weight transfer is observed in both A$_{1g}$ and E$_{2g}$ symmetries and that it is calculated as $S=\int_{\omega_1}^{\omega_2} \! \chi''(\omega) \, \mathrm{d}\omega.$, which is the one expected to be preserved in the Higgs mode scenario \cite{Varma1981}.

There are still aspects of our experimental results which deserve further scrutiny. In particular our results show symmetry dependence of the 'Higgs' modes energies and widths, as well as symmetry dependent spectral weight transfer upon entering the SC state. Such symmetry dependence certainly deserves further theoretical investigation, by including the $\vec{k}$-space distribution of the energy of the SC gap \cite{NB}, and/or the symmetry dependence of the coupling strength. Generally, even if the SC amplitude mode has a full A$_{1g}$ symmetry (for s-wave superconductors), there is no discrepancy for its experimental observation in the higher irreducible representation E$_{2g}$\cite{varmatobepublished}. \\

The observation of the Higgs mode in bulk superconductors may not be the prerogative of 2H-NbSe$_2$. Indeed, in the A15 family, Nb$_3$Sn and V$_3$Si present a particularly intense and narrow SC mode in the $E_g$ symmetry \cite{Hackl1983,Hackl1989}. In both compounds, this SC mode seems coupled to the phonon mode of the same symmetry. Particularly, in Nb$_3$Sn, the spectral weight of the SC mode and the E$_g$ phonon mode is constant at $\pm~3\%$, reminiscent of our results on NbSe$_2$. Interestingly, in Nb$_3$Sn, the pure pair breaking effect occurs in the A$_{1g}$ symmetry, without any spectral weight transfer from a phonon as expected. Even if no pointed out at that time, it is likely that similar mechanism as in 2H-NbSe$_2$ is at play in these A15 compounds. We note that, in V$_3$Si, it is surprising that such large spectral weight transfer occurs whereas the ratio between the E$_g$ phonon mode energy ($\sim 280~cm^{-1}$) and the SC mode energy ($\sim 40~cm^{-1}$) is large, in contradiction with the theoretical prediction \cite{Varma1982}. The fact that the SC modes were not observed in the samples without martensitic transition, the proximity of the superconducting and martensitic transitions in V$_3$Si, the implication of the E$_g$ mode in the tetragonal distortion\cite{Schicktanz} might point to the martensitic instability as an essential ingredient for the observation of the Higgs mode.

In conclusion, we have reported a Raman scattering investigation of the superconducting mode of the charge density wave superconductor 2H-NbSe$_2$. By comparing 2H-NbSe$_2$ with its SC partner 2H-NbS$_2$ which lacks the CDW order, we have shown that the coexisting CDW mode is a requisite to the observation of the SC mode. 2H-NbS$_2$ exhibits only a weak Cooper-pair breaking peak. In addition, we have precisely measured the spectral weight transfer from the amplitude mode of the CDW to the SC mode in 2H-NbSe$_2$ with decreasing temperature and in both A$_{1g}$ and E$_{2g}$ symmetries. The SC mode draws all its Raman intensity from the coexisting CDW mode. Our experimental findings are fully consistent with the scenario of a SC amplitude mode of 'Higgs' type.
Additional evidences might come from the pressure dependency of the SC and the CDW modes with continuous change of their coupling.

\begin{acknowledgments}
The authors would like to especially thank C.M. Varma for very fruitful discussions and suggestions. We acknowledge B. Dou\c{c}ot, M. Cacciari and V. Lahoche for enlighten us about the Standard Model and G. Blumberg, A. Mialitsin, M. Leroux for stimulating discussion on 2H-NbSe$_2$. We acknowledge I. Paul and C. Ciuti for fruitful discussions. We thank S. Bourmand for her help in samples preparation.
\end{acknowledgments}


\begin{references}

\bibitem{Varma2002} C.M. Varma, J. of Low Temp. Phys. {\bf 126}, 901 (2002).%Higgs Boson in Superconductors.
\bibitem{Nambu} Y. Nambu, Physica {\bf 15D}, 147-151 (1985).%Fermion-Boson Relations in BCS-Type Theories

\bibitem{Podolsky} D. Podolsky, A. Auerbach and D. P. Arovas, Phys. Rev. B {\bf 84}, 174522 (2011).%Visibility of the amplitude (Higgs) mode in condensed matter

\bibitem{Anderson} P.W. Anderson, Phys. Rev. {\bf 110}, 827 (1958).%Coherent Excited States in the Theory of Superconductivity: Gauge Invariance and the Meissner Effect
\bibitem{Higgs1964} P. W. Higgs, Phys. Rev. Lett. {\bf 13}, 508 (1964).%BROKEN SYMMETRIES AND THE MASSES OF GAUGE BOSONS
%\bibitem{Englert1964} F. Englert and R. Brout, Phys. Rev. Lett. {\bf 13}, 321 (1964). le mettre? BROKEN SYMMETRY AND THE MASS OF GAUGE VECTOR MESONS
%autre hoggs boson dans autre systems
\bibitem{endres} M. Endres, T. Fukuhara, D. Pekker, M. Cheneau, P. Schaub, Ch. Gross, E. Demler, S. Kuhr and I. Bloch, Nature {\bf 487}, 454-458 (2012).%The 'Higgs' amplitude mode at the two-dimensional superfluid/Mott insulator transition


\bibitem{Ruegg2008} Ch. R\"{u}egg, B. Normand, M. Matsumoto, A. Furrer, D.F. McMorrow, K.W. Kramer, H.U. Gudel, S.N. Gvasaliya, H. Mutka, and M. Boehm, Phys. Rev. Lett. {\bf 100}, 205701 (2008).%Quantum Magnets under Pressure: Controlling Elementary Excitations in TlCuCl$_3$.


%Kramer,K.W./Gudel,H.U./Gvasaliya,S.N./Mutka,H./Boehm,M./

\bibitem{Matsunaga2013}R. Matsunaga,Y. I. Hamada, K. Makise, Y. Uzawa, H. Terai, Z. Wang, and R. Shimano, Phys. Rev. Lett. {\bf 111}, 057002 (2013).

\bibitem{Varma1982} P.B. Littlewood and C.M. Varma, Phys. Rev. B {\bf 26}, 4883 (1982).%Amplitude collective modes in superconductors and their coupling to charge-density waves.

\bibitem{Varma1981} P.B. Littlewood and C.M. Varma, Phys. Rev. Lett. {\bf 47}, 811 (1981).% Gauge-Invariant Theory of the Dynamical Interaction of Charge Density Waves and Superconductivity.


\bibitem{Sooryakumar80} R. Sooryakumar and M.V. Klein, Phy. Rev. Lett. {\bf 45}, 660 (1980).%Raman, Raman Scattering by Superconducting-Gap Excitations and their Coupling to Charge-Density Waves.


\bibitem{Sooryakumar81} R. Sooryakumar and M.V. Klein, Phy. Rev. B {\bf 23}, 3213 (1981).%Raman, Raman scattering from superconducting gap excitations in the presence of a magnetic field.

\bibitem{Balseiro} C.A. Balseiro and L.M. Falicov, Phy. Rev. Lett {\bf 45}, 662 (1980).%theroy mode G NbSe2
\bibitem{Lei} X.L. Lei, C.S. Ting, and J.L. Birman, Phys. Rev. B {\bf30}, 6387 (1984).%theory peak G,  Spectral function of the charge-density-wave (CDW) phonon in an anisotropic CDW superconductor.


\bibitem{Tutto} I. T\"{u}tt\H{o}, and A. Zawadowski, Phys. Rev. B {\bf 45}, 4842 (1992).%theory G peak
%\bibitem{Blumberg} Blumberg, private communication.


\bibitem{Tsang} J.C. Tsang, J.E. Smith, Jr., and M.W. Shafer, Phy. Rev. Lett. {\bf 37}, 1407 (1976). %Raman
% Raman Spectroscopy of Soft Modes at the Charge-Density-Wave Phase Transition in 2H-NbSe$_2$.

\bibitem{Wang} C.S. Wang and J.M. Chen, Sol. State Com. {\bf 14}, 1145 (1974). %Raman, Raman Spectrum of Metallic Layered Compound NbSe$_2$.


\bibitem{Pereira} C.M. Pereira and W.Y. Liang, J. Phys. C: Sol. State Phys. {\bf 15}, L991-L995 (1982). %Raman, Raman studies of the normal phase of 2H-NbSe$_2$. {\it


\bibitem{Wu} Y. Wu, M. An, R. Xiong, J. Shi, and Q.M Zhang, J. Phys. D: Appl. Phys. {\bf 41}, 175408 (2008).%Raman, Raman scattering spectra in the normal phase of 2H-NbSe$_2$.


\bibitem{Mialitsin} A. Mialitsin, J. Phys. and Chem. of Sol. {\bf 72}, 568-571 (2010).%Raman NbSe2, Fano line shape and anti-crossing of Raman active E$_{2g}$ peaks in the charge density wave state of NbSe$_2$.


%mesure Raman NbS2:

\bibitem{Nakashima} S. Nakashima, Y. Tokuda, A, Mitsuishi, R. Aoki, and Y. Hamaue, Sol. State Com. {\bf 42}, 601-604 (1982).%NS2 Raman, Raman Scattering from 2H-NbS$_2$ and Intercalated NbS$_2$.


%\bibitem{McMullan83} McMullan, W.G. \& Irwin, J.C. Raman Scattering from 2H and 3R-NbS$_2$. {\it Sol. State Com.} {\bf 45}, 557-560 (1983). %Raman NbS2


\bibitem{McMullan85} W.G. McMullan and J.C. Irwin, Phys. Stat. Sol. {\bf 129}, 465 (1985). %Raman NbS2, An Interpolytypical Transition in NbS$_2$.

\bibitem{GuillamonS} I. Guillam\'on, H. Suderow, S. Vieira, L. Cario, P. Diener, and P. Rodi\'ere, Phys. Rev. Lett. {\bf 101}, 166407 (2008).%STM NbS2, Superconducting Density of States and Vortex Cores of 2H-NbS$_2$.
\bibitem{kacmarcik} J. Ka\v{c}mar\v{c}\'{i}k, Z. Pribulova, C. Marcenat, T. Klein, P. Rodi$\grave{e}$re, L. Cario, and P. Samuely, Phys. Rev. B {\bf 82}, 014518 (2010). %cp NbS2 NbS2, Specific heat measurements of a superconducting NbS$_2$ single crystal in an external magnetic field: Energy gap structure.


\bibitem{Diener} P. Diener, M. Leroux, L. Cario, T. Klein, and P. Rodi\`{e}re, Phys. Rev. B {\bf 84}, 054531 (2011).%penetration depth,  In-plane magnetic penetration depth in NbS$_2$.




\bibitem{Kiss} T. Kiss, T. Yokoya, A. Chainani, S. Shin, T. Hanaguri, M. Nohara, and H. Takagi, Nature Physics {\bf 3}, 720-725 (2007).%ARPES NBSe2, Charge-order-maximized momentum-dependent
%superconductivity.

\bibitem{GuillamonSe} I. Guillam\`on, H. Suderow, F. Guinea, and S. Vieira, Phys. Rev. B {\bf 77}, 134505 (2008).%STM NbSe2, ntrinsic atomic-scale modulations of the superconducting gap of 2H-NbSe$_2$.




\bibitem{Boaknin} E. Boaknin, M. A. Tanatar, J. Paglione, D. Hawthorn, F. Ronning, R.W. Hill, M. Sutherland, L. Taillefer, J. Sonier, S.M. Hayden, and J.W. Brill, Phys. Rev. Lett. {\bf 90}, 117003 (2003).%Heat Conduction in the Vortex State of NbSe$_2$: Evidence for Multiband Superconductivity.
%« Heat Conduction in the Vortex State of NbSe2 : Evidence for Multiband Superconductivity ». Dans : Phys. Rev. Lett. 90 (11 mar. 2003), p. 117003. doi : 10 . 1103 / %PhysRevLett.90.117003 (cf. p. 38).

\bibitem{Takahashi} private communications from Prof. Takahashi's group, Tohoku University, Sendai.%ARPES NbS2





\bibitem{Borisenko} S. V. Borisenko, A. A. Kordyuk, V. B. Zabolotnyy, D. S. Inosov, D. Evtushinsky, B. B$\ddot{u}$chner, A. N. Yaresko, A. Varykhalov, R. Follath, W. Eberhardt, L. Patthey, and H. Berger, Phys. Rev. Lett. {\bf 102}, 166402 (2009).%arpes%    Two Energy Gaps and Fermi-Surface ''Arcs'' in NbSe$_2$.

\bibitem{Leroux2012} M. Leroux, M. Le Tacon, M. Calandra, L. Cario, M-A. M\'easson, P. Diener, E. Borrissenko, A. Bosak, and P. Rodi\`{e}re, Phys. Rev. B {\bf 86}, 155125 (2012). % X-ray NbS2
%Anharmonic suppression of charge density waves in 2H-NbS2



\bibitem{Fisher1980} W. Fisher and M. Sienko, Inorg. Chem. {\bf 19}, 39-43 (1980).%preparation samples, J. Stoichiometry, Structure, and Physical Properties of Niobium Disulfide.


\bibitem{sup1}See Supplemental Material at [] for the evaluation of the laser heating. 



\bibitem{Abrikosov1961} A. A. Abrikosov and L. A. Falkovskii, Zh. Eksp. Teor. Fiz. {\bf 40}, 262 (1961).
\bibitem{Abrikosov1973} A.A. Abrikosov and V.M. Genkin, Zh. Eksp. Teor. Fiz. {\bf 65}, 842 (1973). (Engl. Transl. Sov. Phys. JETP {\bf 38}, 417 (1974)).
\bibitem{Klein1984} M.V. Klein and S.B. Dierker, Phys. Rev. B {\bf 29}, 4976 (1984).
\bibitem{DevereauxRMP} T. P. Devereaux, and R. Hackl, Rev. of Mod. Phys. {\bf 79}, 175 (2007).%R. Inelastic light scattering from correlated electrons.

%A.A. Abrikosov and V.M. Genkin, Zh. Eksp. Teor. Fiz. 65 842 (1973)
%(Engl. Transl. Sov. Phys. JETP 38 417 (1974)) (in addition to
%Abrikosov and Falkovskii) and M.V. Klein and S.B. Dierker PRB 29, 4976
%(1984).


\bibitem{sup2} See Supplemental Material at [] for the pure A$_{1g}$ Raman signals. 


\bibitem{NB} 2H-NbSe$_{2}$ is believed to be a multi-gap superconductor. No feature related to the presence of the small gap in the Raman spectra of 2H-NbSe$_2$ has been detected.


\bibitem{varmatobepublished} D. Pekker and C.M. Varma, to be published.

\bibitem{Hackl1983}R. Hackl, R. Kaiser and S. Schicktanz, J. Phys. C: Solid State Phys. {\bf 16}, 1729 (1983).%Gap mode, superconducting gap and phonon mode in V3Si and Nb3Sn

\bibitem{Hackl1989}R. Hackl, R. Kaiser and W. Glasersi, Physica C {\bf 162-164}, 431 (1989).%ELECTRONIC AND PHONON-MEDIATED GAP MODES IN A15 COMPOUNDS

\bibitem{Schicktanz} S. Schicktanz, R. Kaiser, E. Schneider, and W. Gl\"{a}ser, Phys. Rev. B {\bf 22}, 2386 (1980).
%Raman studies of 215 compounds













\end{references}
\end{document}